\documentclass[groupedaddress]{revtex4}

\usepackage{graphicx}

\newlength\NSIZE
\begin{document}

\title{MovieMaker: A Parallel Movie-Making Software for Large Scale Simulations}

\author{Hitoshi Uehara}
\author{Shintaro Kawahara}
\author{Nobuaki Ohno}
\author{Mikito Furuichi}
\author{Fumiaki Araki}
\author{Akira Kageyama}

\affiliation{Earth Simulator Center, \\
Japan Agency for Marine-Earth Science and Technology, \\
3173-25 Showa-machi, Kanazawa-ku, Yokohama, 236-0001, Japan}

\begin{abstract}
We have developed a parallel rendering software for scientific visualization of large-scale, three-dimensional, time development simulations.
The goal of this software, MovieMaker, is to generate a movie, or a series  of visualization images from totally one TB-scale data within one night (or less than 12 hours).
The isocontouring, volume rendering, and streamlines are implemented.
MovieMaker is a parallel program for the shared memory architecture with dynamic load balancing and overlapped disk I/O.
\end{abstract}

\maketitle

\section{Introduction}

The output data size of computer simulation is rapidly growing in accordance with the development of high performance computers.
On Earth Simulator, it is not seldom that output data reaches more than 1 GB per one time step for a single variable. Since a typical simulation may produce the output data for several variables for hundreds or sometimes thousands time steps, the whole data set by one simulation job can reach more than 1 TB in total.

Three dimensional, time development data should be visualized by a series of images, or a movie.
However, making a scientific visualization movie from 1 TB data in a reasonable time is beyond the ability of commercially available software today.
 
Therefore, we have decided to develop such a software by ourselves.
The goal of the system, ``MovieMaker", is to enable us to produce a movie from one TB simulation data within 12 hours, or one night\cite{B:goudou2005}.
When the whole data set consists of 1,024 time steps with 1 GB each, MovieMaker should be able to apply various visualization processes for each time step within about 40 seconds.

MovieMaker can handle two different kinds of visualization methods, the polygon-based method and the volume-based method.
This is useful when one should visualize a mixed set of scalar and vector fields; for instance, the volume rendering for a pressure field and the streamlines for a flow vector field.
This can be contrasted with most of other visualization tools developed so far \cite{B:YU,B:MA,B:PISA}.
Another feature of MovieMaker is its simple disk I/O.
Yu et al.\cite{B:YU} pointed out the disk I/O cost in the parallel rendering system.
They solved this problem by using the parallel disk I/O.
On the other hand, we use the standard non-parallel disk I/O.

\section{Development of MovieMaker}

\begin{figure}
  \centering
  \includegraphics[width=0.8\textwidth]{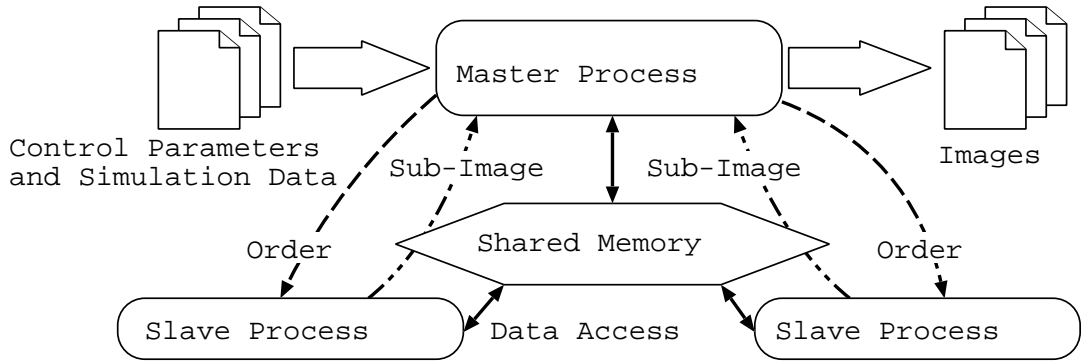}
  \caption{MovieMaker. Master/slave model for the parallel processing \protect\\
on the shared memory architecture is adopted.}
  \label{F:arch}
\end{figure}

We have designed MovieMaker as a master/slave parallel rendering program for the shared-memory architecture.
As shown in Figure \ref{F:arch}, the master process and slave processes share the simulation data stored in the shared memory area. 
The master process performs the following multi-tasks;
(i) to read a configuration file,
(ii) to read the simulation data into shared-memory area;
and (iii) to control the slave processes keeping a good load balance.
Slave processes perform rendering tasks following commands sent from the master process and then return partial-images back to the master via shared memory.
Interprocess communications (IPC) are performed with Message Passing Interface (MPI).
We have achieved the dynamic load balancing in MovieMaker by an active monitoring and dynamic control of the slave processes.

\begin{figure}
  \centering
  \includegraphics[width=0.8\textwidth]{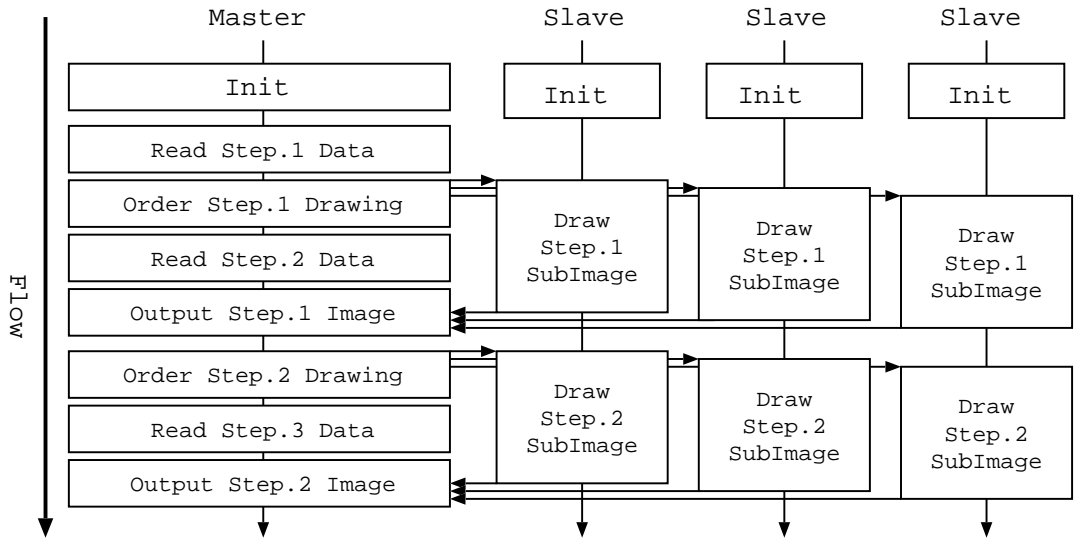}
  \caption{Flowchart of MovieMaker. Overlapping of data read step\protect\\
and rendering step is achieved.}
  \label{F:flow}
\end{figure}

In addition to the dynamic load balancing, 
we have also implemented task overlapping for efficient parallel rendering.
Figure \ref{F:flow} shows a flowchart of MovieMaker.
While the slave processes are working for rendering tasks for one specific time step of the simulation data, the master process reads data of the next time step.
Since it takes about a second to read a GB scale data for each time step from the hard disk drive, the total read time for the whole movie data may reach about an hour.
We could hide this read time in the rendering time of MovieMaker by applying the task overlapping method.

MovieMaker is implemented in C++ with OpenGL. 
OpenGL is used for the off screen rendering.
The present version of MovieMaker has three visualization methods:
1) Volume-rendering; 2) isocontouring; and 3) streamlines.
The volume rendering and the isocontouring are implemented based on basic algorithms of the ray casting and the marching-cubes, respectively.
These three visualization methods can be used in juxtaposition.
Fine-tuning of each visualization method is possible by controlling visualization parameters specified by the configuration file.

MovieMaker, in the present version, adopts input data defined in the Cartesian geometry. 
The grid system should be rectilinear, i.e., uniform and/or non-uniform in each direction
of $x$, $y$, and $z$. The maximum grid number is constrained only by the shared memory size.

\begin{figure}
\centering
\includegraphics[width=\NSIZE]{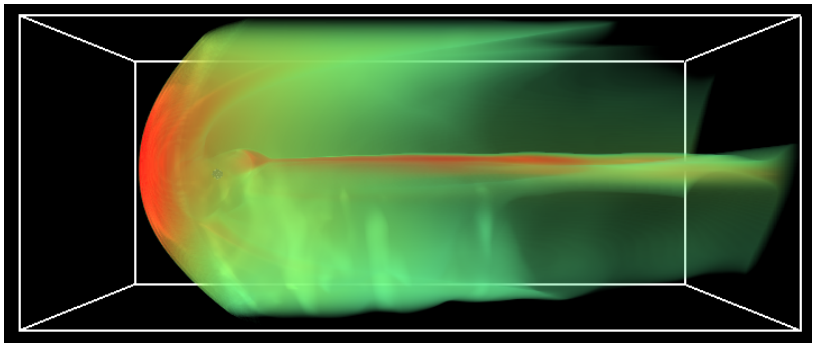}
\caption{A sample image generated by MovieMaker using the volume rendering.}
\label{F:vr}
\end{figure}
\begin{figure}
\centering
\includegraphics[width=0.6\NSIZE]{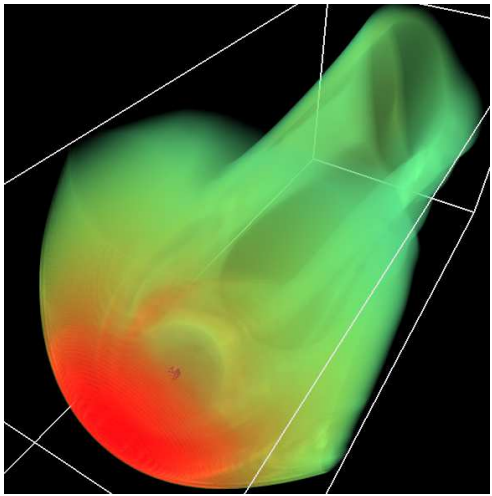}
\caption{Same as Figure \ref{F:vr} viewed from different angle.}
\label{F:vr2}
\end{figure}
\begin{figure}
\centering
\includegraphics[width=\NSIZE]{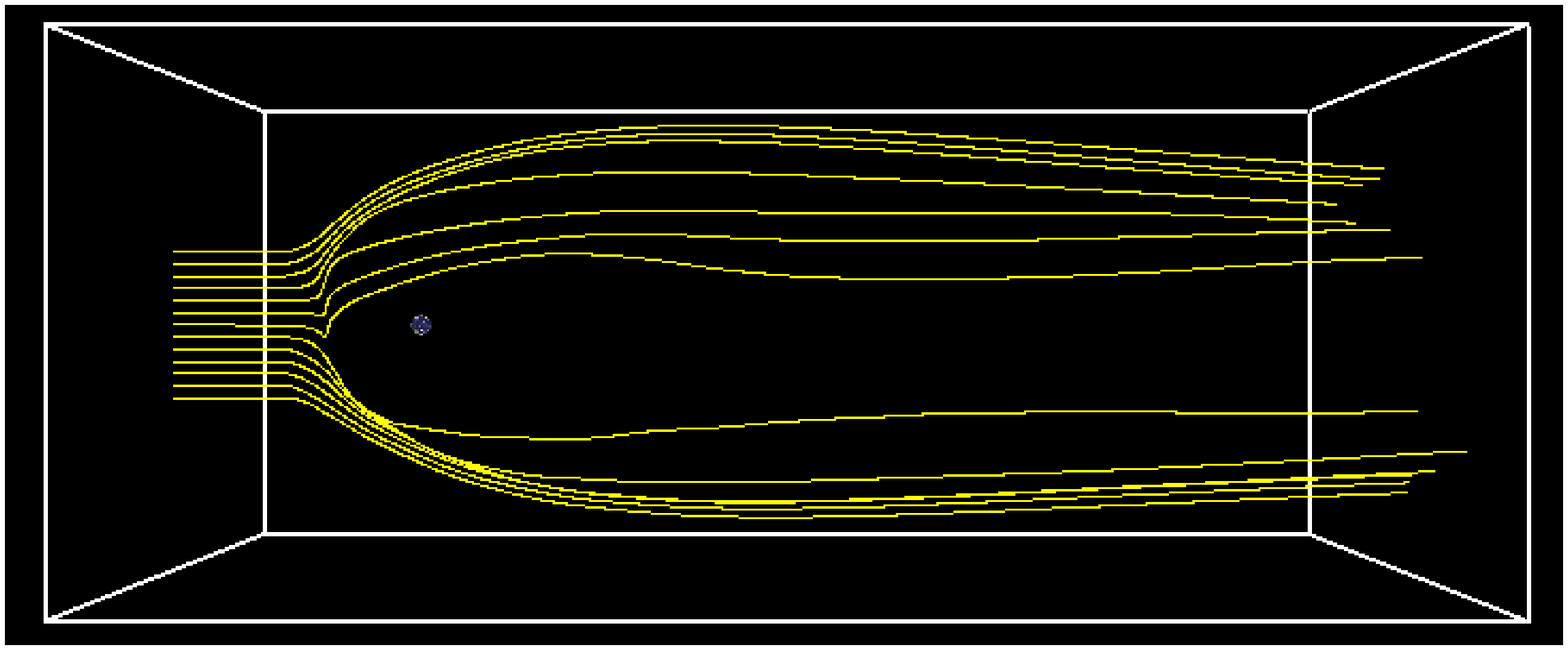}
\caption{A sample image generated by MovieMaker using streamlines.}
\label{F:stl}
\end{figure} 

In our computational environment with SGI Onyx 3800 (12CPU, 24GB memory; 11 MPI processes) , MovieMaker has achieved the target performance, i.e., processing one GB within 40 seconds.
A performance test was applied to a geodynamo simulation data with the grid size 640 $\times$ 640 $\times$ 640 of 4-byte floating point. The average processing time to generate a single image by the volume rendering, isocontouring, and streamlines are, 40 seconds, 33 seconds, and 10 seconds, respectively.

Figure  \ref{F:vr} and \ref{F:vr2} show sample images of the volume rendering visualization applied to the temperature distribution of an MHD simulation of the Earth's magnetosphere.
High temperature regions are emphasized.
The grid-size of the data is 502 $\times$ 202 $\times$ 202, with the 4-byte floating-point data. The data size for one step is about 80 MB.

Figure  \ref{F:stl} is another sample image generated by MovieMaker.
This image shows the velocity of solar wind in the simulation of the magnetosphere, using the streamlines.
The bow shock is clearly seen.

\section{Conclusion}

We have developed MovieMaker, a parallel movie-making software for scientific visualization of large-scale simulations. The main goal of MovieMaker is to make a movie from a TB data within one night (or less than 12 hours), and this goal has been successfully achieved for the basic visualization methods including the volume rendering, isocontouring, and streamlines.

\end{document}